\documentclass{emulateapj}

\def\xmm{{\it XMM-Newton}}

\def\eps{{erg s$^{-1}$}}
\def\pcm{{cm$^{-2}$}}
\def\NH{$N_{\rm H}$}

\slugcomment{To Apper  in  {\it The Astrophysical Journal}}
 
\shorttitle{Soft X-ray AGN}

\shortauthors{Terashima et al.}

\begin{document}

\title{A Candidate Active Galactic Nucleus with a Pure Soft Thermal X-ray Spectrum\footnote
{
Based on observations obtained with {\it XMM-Newton}, an ESA science mission with instruments and contributions directly funded by ESA Member States and NASA
}
}
\author{Yuichi Terashima\altaffilmark{1}, 
Naoya Kamizasa\altaffilmark{1}, 
Hisamitsu Awaki\altaffilmark{1}, 
Aya Kubota\altaffilmark{2} , 
and Yoshihiro Ueda\altaffilmark{3}
}


\altaffiltext{1}{Department of Physics, Ehime University, Matsuyama, Ehime 790-8577, Japan}

\altaffiltext{2}{
Department of Electronic Information Systems, Shibaura Institute of Technology, Saitama, Saitama 337-8570, Japan}

\altaffiltext{3}{Department of Astronomy, Kyoto University, Kyoto 606-8502, Japan}

\date{}

\begin{abstract}


We report the discovery of a candidate active galactic nucleus (AGN), 2XMM~J123103.2+110648 at $z = 0.13$, with an X-ray spectrum 
represented purely by soft thermal emission reminiscent of Galactic black hole (BH) binaries in the disk-dominated state. This object was found
in the second {\xmm} serendipitous source catalogue as a highly variable X-ray source. 
In three separate observations, its X-ray spectrum can be represented either by 
a multicolor disk blackbody model with an inner temperature of $kT_{\rm in}\approx 0.16-0.21$ keV 
or a Wien spectrum Comptonized by an optically thick plasma with $kT \approx 0.14-0.18$ keV.
The soft X-ray luminosity in the 0.5--2 keV band is estimated to be (1.6-3.8)$\times 10^{42}$ {\eps}.
Hard emission above $\sim$ 2 keV is not detected. The ratio of the soft to hard 
emission is the strongest among AGNs observed thus far. 
Spectra selected in high/low flux time intervals are examined in order to study spectral variability.
In the second observation with the highest signal-to-noise ratio, 
the low energy (below 0.7 keV) spectral regime flattens when the flux is high, while the shape of the high energy part (1--1.7 keV)
remains unchanged.  This behavior is qualitatively consistent with being caused by strong Comptonization.
Both the strong soft excess and spectral change consistent with Comptonization in the X-ray spectrum imply that the Eddington
ratio is large, which requires a small BH mass (smaller than $\sim10^5M_{\odot}$).

\end{abstract}
\keywords{galaxies: active, X-rays: galaxies}

\section{Introduction}

It is well known that stellar mass black holes (BHs) show state transitions depending upon mass accretion 
rate and that X-ray spectra show different characteristics in each state.
Such distinct states, however,  are much less understood in active galactic nuclei (AGN), which contain super massive BHs. The typical X-ray spectrum of unobscured
AGN consists of several components including a power law with an exponential cutoff at an energy above $\sim$100 keV, a ``soft excess'', and emission reprocessed
by matter around the central X-ray source. Spectra dominated by a cutoff  power law are reminiscent of spectra of Galactic BH binaries (BHBs) in their low/hard state. 
In addition to the hard power law component, many AGN show soft excess emission, 
which is approximated by a blackbody or a multicolor disk  (MCD) blackbody (Mitsuda et al 1984)
with a temperature $kT \approx 0.1-0.2$ keV. 
In a subset of AGN, the narrow-line Seyfert 1s (NLS1s), the soft excess component
is often more prominent.
While the form of this component is apparently similar to disk blackbody
emission in BHBs in their high/soft (HS) state (Makishima et al. 1986), there are difficulties in ascribing its origin to be the same as the disk emission, and it is believed that most AGN are 
not in the HS state at least in terms of their X-ray spectra.
On the other hand, there is also an argument that properties of X-ray variability such as power spectral 
densities in AGN
are more like the HS state in BHBs (McHardy et al. 2006). Thus, classification of spectral states in AGN is still in dispute.

AGN exhibiting strong soft excess emission are of great interest in studying 
many aspects of BHs and accretion such as the origin of
the soft excess and spectral states in AGN. Furthermore, it is easier to measure the shape of the soft excess without dilution by
the hard power law component in such AGN. In this paper, we report 
discovery of an AGN with an X-ray spectrum purely explained by soft emission. The quoted
errors are at a 90\% 
confidence level for one parameter of interest.
We assume a cosmology of $H_0 = 70$ km s$^{-1}$ Mpc$^{-1}$, $\Omega_{\rm M}$ = 0.3, and $\Omega_{\Lambda} $= 0.7.



\begin{table*}[ht]
\begin{center}
\caption{Observation Log}
\begin{tabular}{ccccccc}
\hline
\hline
No  & Observation ID & Observation Start Date  & \multicolumn{3}{c}{Exposure Time (ks)} \\
			& 					&  & EPIC-PN   & EPIC-MOS1 & EPIC-MOS2\\
\hline
1 & 0145800101	& 2003 July 13		& 46.3		& 54.8 	& 56.4\\ 
2 & 0306630101	& 2005 Dec. 13		& 54.6		& ...		& 66.4\\
3 & 0306630201	& 2005 Dec. 17		& 80.4		& ...		& 91.5\\
\hline
\end{tabular}
\end{center}
\end{table*}

\vspace{5mm}

\section{Observation and Data Analysis}

The highly variable object with a soft X-ray spectrum 2XMM~J123103.2+110648  (hereafter 2XMM~J1231) was discovered
in the course of our systematic study of variable AGN (Kamizasa et al. 2012)
in the second {\xmm} serendipitous source catalogue (Watson et al. 2009).
A likely candidate of the optical counterpart is 
SDSS~J123103.24+110648.5 clasified as a galaxy, which is the only optical source within the error circle of the X-ray position
and is slightly extended in optical images. The optical extent of this object is much smaller than the position accuracy
of {\xmm}, and it is not clear whether the X-ray source coincides with the nucleus of the galaxy.
No optical spectra are available for this source, and 
 its photometric redshift is estimated to be
$z = 0.13$ with an error of 0.05 based on multiband photometry data of the Sloan Digital Sky Survey (SDSS).
We assume this redshift throughout this paper unless otherwise noted.

2XMM~J1231 is serendipitously detected in  {\xmm} observations of the quasar LBQS 1228+1116,
and was observed with {\xmm} three times (ObsID 0145800101, 0306630101, and 0306630201).
These three observations are referred to as observation 1, 2, and 3, respectively.
The observation log is shown in Table 1.
2XMM~J1231 was detected in the field of view of EPIC-PN and EPIC-MOS2 in all observations.
The CCD of EPIC-MOS1 at the
source position was disabled due to damage in observations 2 and 3, and therefore EPIC-MOS1
data are available only for observation 1.
We made light curves and spectra from each data set as follows. We reprocessed the data 
with the Science Analysis Software (SAS) version 11 and calibration data as of 2011 March.
The light curves and spectra were extracted from a circular region with a radius of 20$^{\prime\prime}$
centered on
the source. Light curves from a source free region in the energy band above 10 keV were examined and
high background time intervals were discarded. The effective exposure times 
after data screening are shown in Table 1.
Background spectra were made from a nearby source free region and subtracted from the source spectrum
during spectral analysis. Response files were made by using {\tt rmfgen} and {\tt arfgen} in the SAS package.
The extracted spectra were binned to contain at least 20 counts per bin so that chi-square
fitting can be used. We fitted the spectra obtained by EPIN-PN and EPIC-MOS simultaneously. 


We examined the counts detected in various energy bands, and found that this object has a very soft
X-ray spectrum. In the hard energy band (2--7 keV observed frame), no significant signal was detected. The detected number of PN 
counts  in the source extraction region were 44, 66, and 90 for the three observations, respectively,
which are consistent with the counts expected from background (52, 65, and 86 counts). The number of PN counts in 1.7--2 keV 
are also low (8, 18, and 17 counts for the three observations, respectively) compared to the expected
background (9, 11, and 14 counts). By contrast, significant signal is clearly seen below 1.7 keV.



We also examined UV images obtained with the OM. Images through the three filters UVW1, UVM2, and UVW2, with effective wavelengths of 2910 \AA , 2310 \AA , and 2120 \AA, respectively, 
were obtained for observations 1 and 3, while only UVM2 and UVW2 were used in observation 2.
Multiple exposures were combined to single images for each filter, and source detections were performed using the
{\tt omichain} task in the SAS. A possible UV counterpart of 2XMM~J1231 is found only in UVW1 images. The fluxes measured 
from observations 1 and 3 are $(1.6\pm0.1) \times 10^{-17}$ and 
$(2.1\pm0.9) \times10^{-17}$ erg s$^{-1}$ cm$^{-2}$ \AA$^{-1}$, respectively.  No variability between the two observations
is seen. We use the former flux with better accuracy below.

\begin{figure}[ht]
\begin{center}
\includegraphics[angle=-90,scale=0.36]{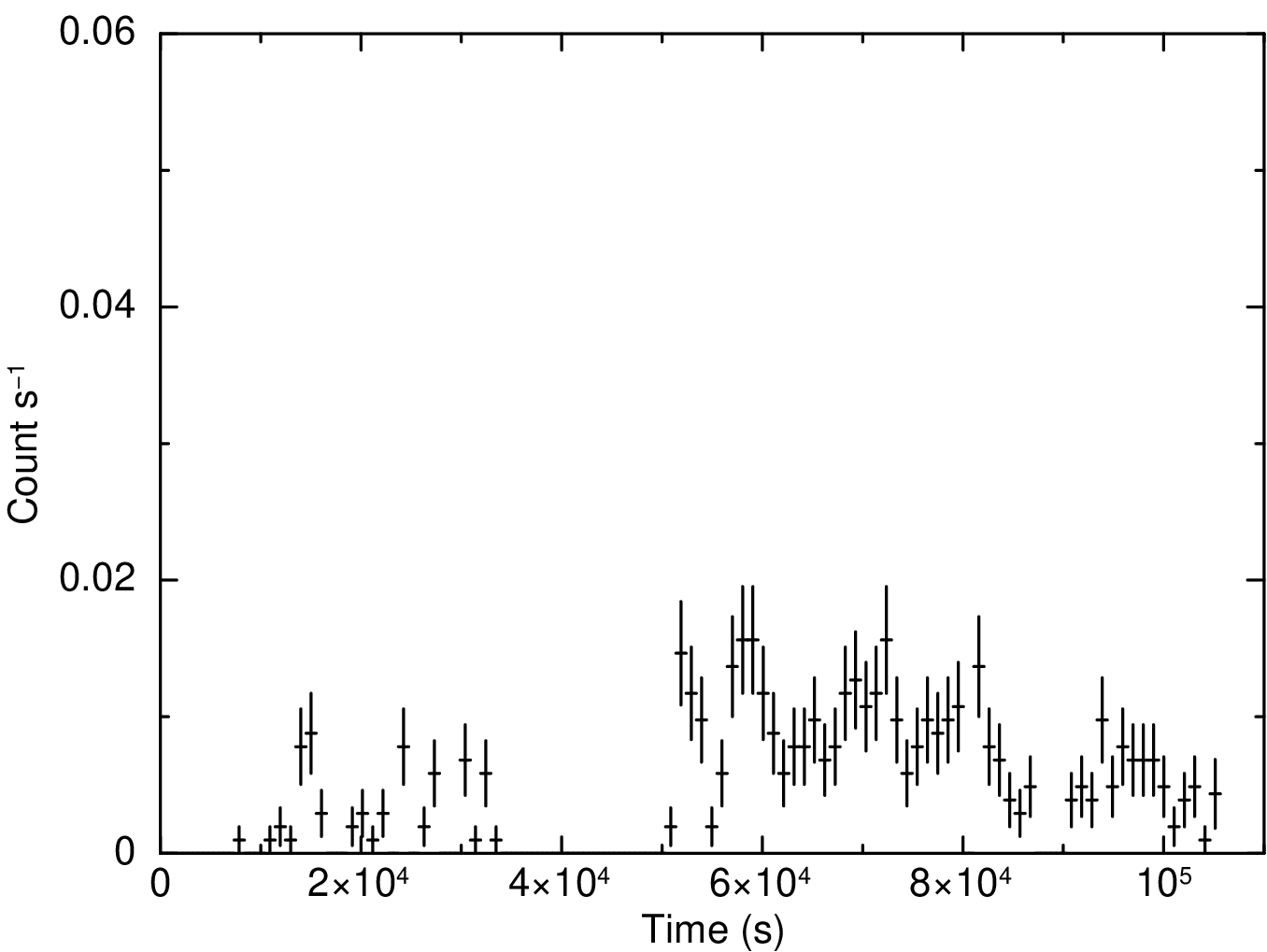}
\includegraphics[angle=-90,scale=0.36]{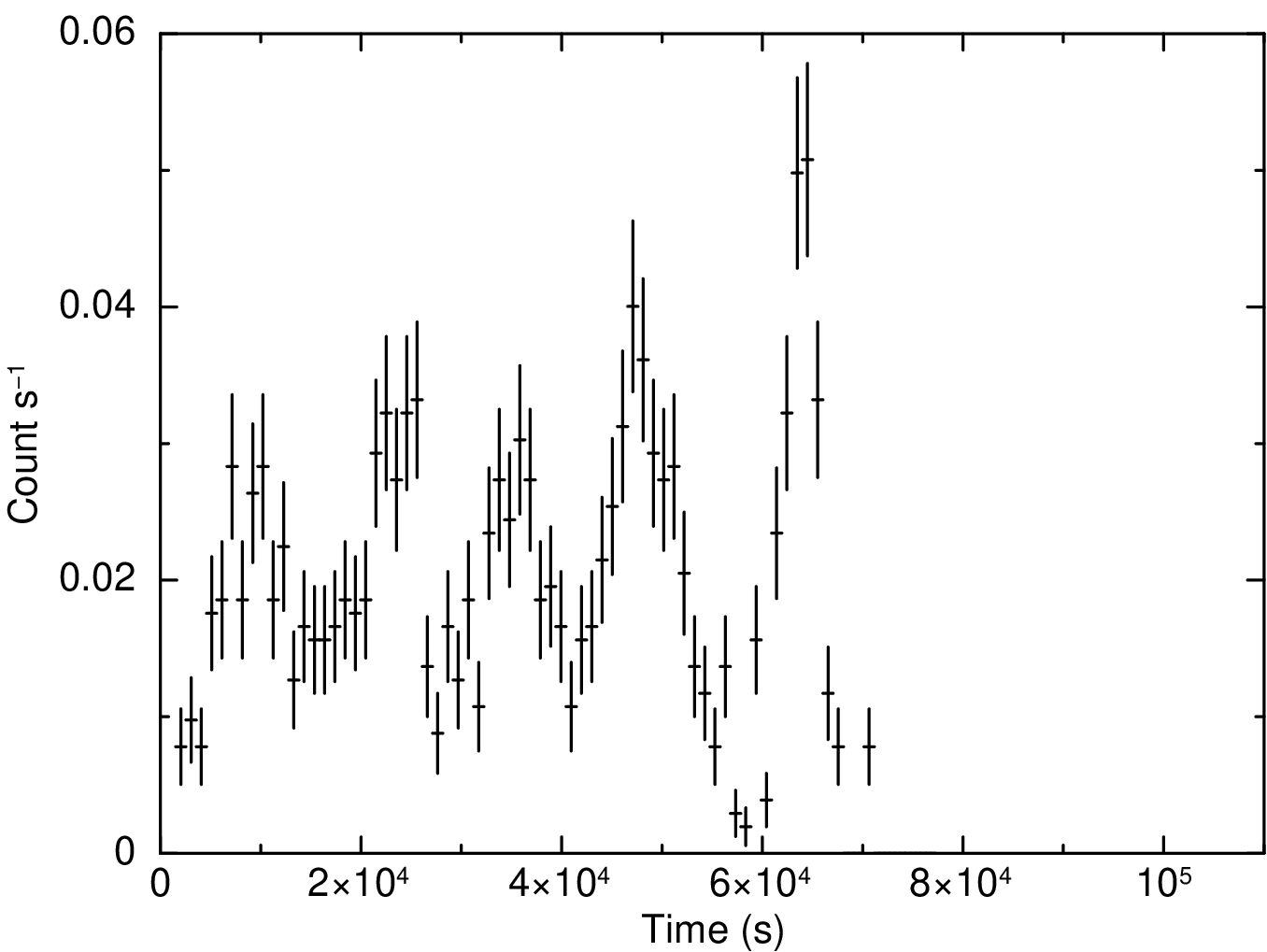}
\includegraphics[angle=-90,scale=0.36]{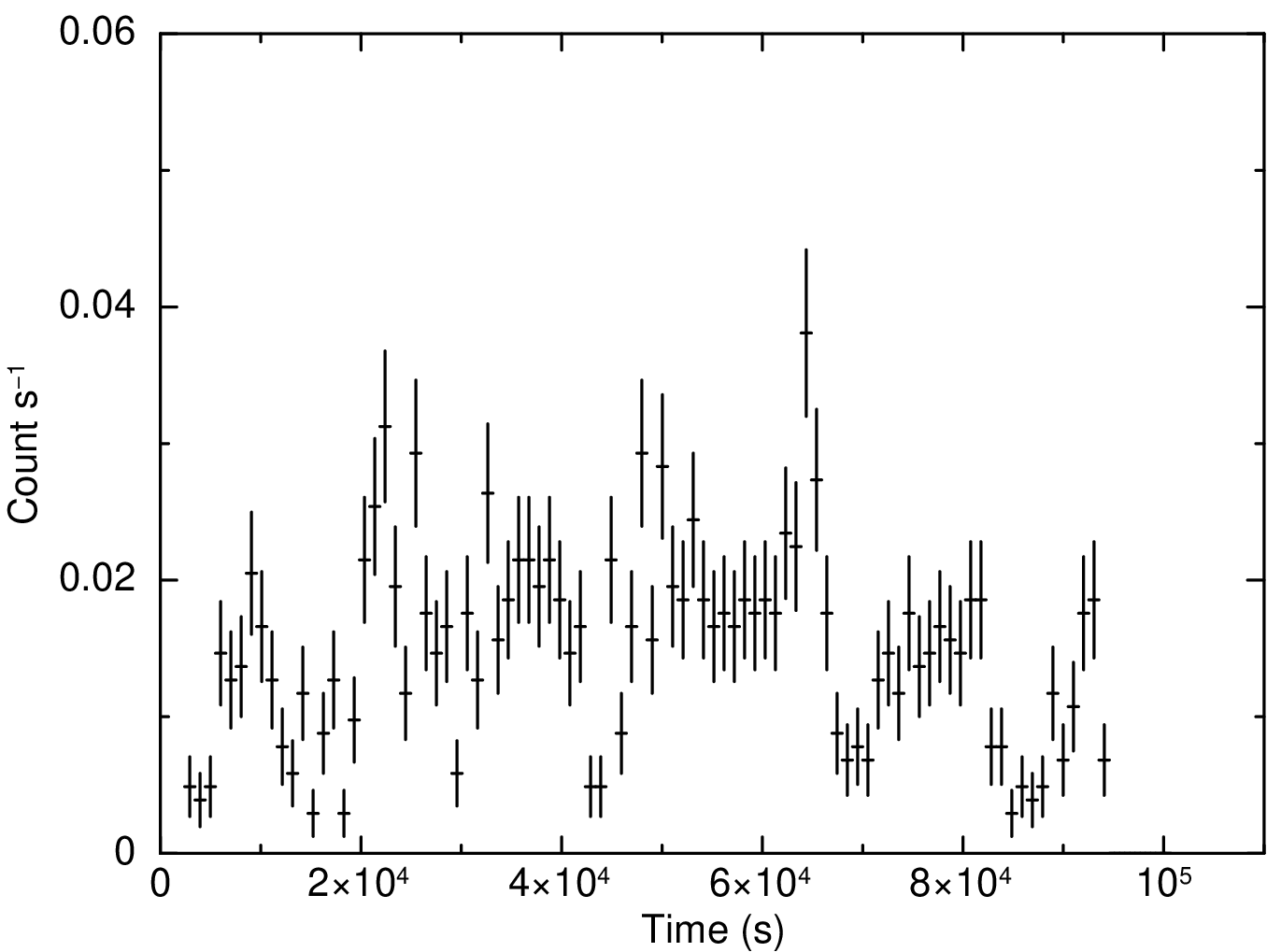}
\caption{Light curves of 
2XMM~J123103.2+110646
obtained with EPIC-PN in the 0.44--1.77 keV band (observed frame) for observations 1 to 3 from top
to bottom.
Bin size is 1024 s.}
\end{center}
\end{figure}

\begin{figure}[ht]
\begin{center}
\includegraphics[angle=-90,scale=0.36]{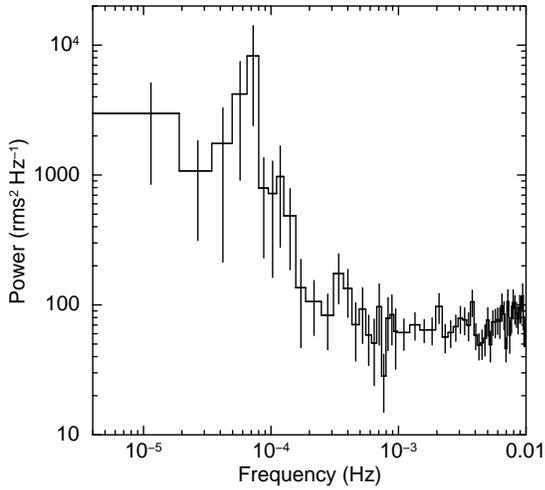}
\caption{
Power spectrum of a light curve in 0.3--1.7 keV for observation 2.
}
\end{center}
\end{figure}

\begin{figure}[ht]
\begin{center}
\includegraphics[angle=-90,scale=0.36]{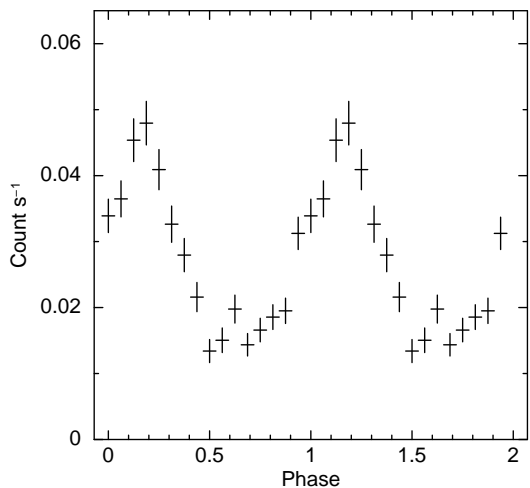}
\caption{
Folded light curve in 0.3--1.7 keV for observation 2. The folded period is 
14000 s and the origin of the phase is 2005 Dec. 13 10:32:42 (UT).
}
\end{center}
\end{figure}

\vspace{2cm}

\section{Results}

\subsection{Light Curves}

Light curves of 2XMM~J1231 in the 0.44--1.77 keV band (observed frame) or the 0.5--2 keV band (source rest frame) 
binned to 1024 s are shown in Fig. 1. 
The X-ray flux is highly variable and changed by a factor of more than three
on timescales of $\sim1000$ s. 
We calculated the normalized excess variance and its error using the expression 
given in Vaughan et al. (2003) from these light curves as $0.47\pm0.04, 0.45\pm0.02$, and $0.41\pm0.02$, 
for observations 1, 2, and 3, respectively.



Possible quasi-periodic modulation with a period of $\sim15000$ s
is seen in the light curve for observation 2. Hints of modulations with a similar period are also seen in light curves for
observations 1 and 3. Since the possible periodicity is most 
clearly seen in the light curve of the second observation,
we made a power spectrum 
using a light curve in the 0.3--1.7 keV for observation 2 obtained with EPIC-PN.
The power spectrum, as shown in Fig. 2, shows a peak around 
$7\times10^{-5}$ Hz, though errors are large. We folded  the light curve over a range
of trial periods and calculated  chi-square values based on a hypothesis of no variation.
The period providing the maximum chi-square was determined
to be $1.40\times10^4$ s, where $\chi^2 = 322.9$ was obtained for a light curve
folded at this period with 20 bins.
The light curve folded at this period for observation 2
is shown in Fig. 3.

\begin{table*}[ht]
\begin{center}
\caption{Results of Spectral Fits}
\begin{tabular}{ccccccc}
\hline
\hline
Model$^{\rm a}$ & Parameters & \multicolumn{3}{c}{Value}\\
 & & Observation 1  & Observation 2 & Observation 3\\  
\hline
Blackbody & Absorption$^{\rm b}$ ($10^{20}$ cm$^{-2}$) & 0.0 ($<2.2$) & 0.0 ($<1.1$) & 0.0 ($<6.2$)\\
		& $kT$ (keV) & $0.13\pm0.01$ & $0.15\pm0.01$  & $0.14\pm0.01$\\
		& Normalization$^{\rm c, d}$ & $1.2\times10^{-6}$ & $2.4\times10^{-6}$  & $1.6\times10^{-6}$\\
		& $\chi^2$ (dof)  & 30.2 (32)  & 72.3 (57)  & 62.1 (61)\\
MCD		& Absorption$^{\rm b}$ ($10^{20}$ cm$^{-2}$) &  0 ($<4$) & 0 ($<4$)  & $7^{+3}_{-6}$\\ 
		& $kT$ (keV)& $0.18^{+0.01}_{-0.02}$ & $0.21^{+0.01}_{-0.03}$   & $0.16\pm0.02$\\
		& Normalization$^{\rm e}$ & 8.3 & 7.8 & 35 \\
		& $\chi^2$ (dof)  &  22.0 (32) & 62.7 (57)   & 59.4 (61)\\
CompTT$^{\rm f}$   & Absorption$^{\rm b}$ ($10^{20}$ cm$^{-2}$) & 0 ($<6$) & 0 ($<4$) & 7 ($<10$)\\
		& $kT$ (keV) & $0.18^{+0.02}_{-0.03}$  & $0.18^{+0.01}_{-0.02}$  & $0.14^{+0.02}_{-0.01}$ \\
		& $\tau$ & $21^{+9}_{-4}$ & $30^{+20}_{-10}$  & $30^{+170\rm ,g}_{-20}$\\
		& Normalization & $2.3\times10^{-2}$ & $1.3\times10^{-2}$ & $3.4\times10^{-2}$\\
		& $\chi^2$ (dof)  &  20.2 (31) & 62.3 (56)  & 60.2 (60)\\
\hline
\end{tabular}
\end{center}
\tablecomments{
$^{\rm a}$ Redshift of the source is assumed to be $z=0.13$.\\
$^{\rm b}$ Absorption column density intrinsic to the source. The Galactic absorption is assumed to be {\NH} = 2.3$\times10^{20}$ {\pcm}.\\
$^{\rm c}$ Normalization determined by the EPIC-PN data.\\
$^{\rm d}$
$L_{39}/D_{10}^2$, where $L_{39}$ is the source luminosity in units of $10^{39}$ erg s$^{-1}$, and $D_{10}$ is the distance to the source in units of 10 kpc.\\
$^{\rm e}$
$ R_{\rm in}/D_{10}^2 \cos \theta$, where $R_{\rm in}$ is the apparent inner disk radius in units of km, and $D_{10}$ is the distance to the source in units of 10 kpc, $\theta$ the inclination angle of the disk.\\
$^{\rm f}$ The spectrum of the seed photons is assumed to be a Wien law with a temperature of  
10 eV. Disk geometry of Comptonizing plasma is assumed.\\
$^{\rm g}$ Pegged at the largest value allowed in the fit.
}
\end{table*}

\begin{table*}[ht]
\begin{center}
\caption{Spectral Parameters for High and Low Flux States.}
\begin{tabular}{cccccc}
\hline
\hline
Model$^{\rm a}$ & Parameters & \multicolumn{2}{c}{Observation 2} \\
 & & High Flux & Low flux \\
\hline
MCD		& Absorption$^{\rm b}$ ($10^{20}$ cm$^{-2}$) & $13\pm8$ & 0 ($<4$)  \\
		& $kT$ (keV)& $0.18\pm0.03$  & $0.18^{+0.01}_{-0.02}$  \\ 
		& Normalization$^{\rm c, d}$ & 54.8  & 12.7  \\
		& $L_{0.5-2}^{\rm e}$ & $8.9\times10^{42}$ & $2.7\times10^{42}$  \\
		& $\chi^2$ (dof)  &  32.1 (32)  & 65.0 (55)  \\ 
CompTT$^{\rm f}$ 	& Absorption$^{\rm b}$ ($10^{20}$ cm$^{-2}$) & $4 (<10) $ & 2 ($<10$)  \\
		& $kT$ (keV) & $0.16\pm0.02$ &  $0.18^{+0.02}_{-0.03}$  \\  
		& $\tau$ & $200^{\rm g} (>30)$    & $20^{+20}_{-10}$  \\ 
		& Normalization & $3.6\times10^{-3}$  & $6.6\times10^{-2}$  \\  
		& $L_{0.5-2}^{\rm e}$ & $7.0\times10^{42}$ & $2.9\times10^{42}$  \\  
		& $\chi^2$ (dof)  & 29.3 (31) & 62.9 (54)  \\ 
\hline
\end{tabular}
\end{center}
\tablecomments{
$^{\rm a}$ Redshift of the source is assumed to be $z=0.13$.\\
$^{\rm b}$ Absorption column density intrinsic to the source. The Galactic absorption is assumed to be {\NH} = 2.3$\times10^{20}$ {\pcm}.\\
$^{\rm c}$ Normalization determined by the EPIC-PN data.\\
$^{\rm d}$
$ R_{\rm in}/D_{10}^2 \cos \theta$, where $R_{\rm in}$ is the apparent inner disk radius in units of km, $D_{10}$ is the distance to the source in units of 10 kpc, 
and $\theta$ is the inclination angle of the disk.\\
$^{\rm e}$ Luminosity in 0.5--2 keV corrected for absorption in units of erg s$^{-1}$
determined by the EPIC-PN data.\\
$^{\rm f}$ The spectrum of the seed photons is assumed to be Wien law with a temperature of  
10 eV. Disk geometry of Comptonizing plasma is assumed.\\
$^{\rm g}$ Pegged at the largest value allowed in the fit.
}
\end{table*}

\subsection{Spectra}

We attempted to fit the spectra with various continuum models. The spectral fits were performed by using XSPEC version 12.6.0q.
In all the spectral fits, the Galactic absorption derived from 21 cm observations 
(Kalberla et al. 2005)  {\NH} = $2.3\times10^{20}$ {\pcm} was applied
by using the phabs model in XSPEC. The absorption column density intrinsic to the source was treated as a free parameter
and the zphabs model was used with a fixed redshift $z = 0.13$.
We did not select  time intervals common to EPIC-PN and EPIC-MOS2 to achieve better photon statistics. 
Since the source fluxes measured by the two instruments
could be different, the normalization values for EPIC-PN and EPIC-MOS2 spectra were treated as independent parameters.

We first applied a simple power law model to the spectra.
This fit was acceptable only for observation 1 and the best-fit photon index
was extremely steep (4.8). 
The residuals seen in the spectra of observation 2 and 3
clearly show a convex shape suggesting that the spectrum is curved. 
So, a blackbody model was examined and was found to provide better fits.  The slight concave residuals
seen in the spectra of observations 1 and 2, however, show that this model is too curved.
Next we tried to fit the spectra with a thermally Comptonized disk model 
(compTT model in XSPEC, Titarchuk 1994) and an MCD model, both of which are often used to
represent spectra of Galactic BHBs.
In the compTT model, the seed photons were assumed to be a Wien law with a temperature of 10 eV. A disk geometry is assumed for the Comptonizing plasma. As for the MCD model, we modified the diskbb model in XSPEC to properly treat the redshift of the emitter.
Both compTT and MCD models provided good
fits to  the observed spectra.
The spectra fitted with an MCD model are shown in Fig. 4.
The results of the spectral fits are summarized 
in Table 2. 
The absorption-corrected luminosity in the 0.5--2 keV band (source rest frame)
obtained with EPIC-PN
 is  $1.6\times10^{42}$,  $3.8\times10^{42}$, and
 $3.8\times10^{42}$ erg s$^{-1}$ for the three observations, where
the MCD and compTT models provided similar luminosities.


In order to quantify spectral variability, we divided the data into high and low flux intervals.
Among the three observations, observation 2, which shows the largest amplitude variability, displays significant spectral
change.
0.02 counts s$^{-1}$ per PN in the 0.5--1.7 keV band (observed frame)  was used as a boundary between the high/low fluxes.
The compTT and MCD models, which provided good empirical descriptions of the data, were fitted to the
spectra for the two flux states. Both models describe the observed spectra. The spectra along with the best-fit
MCD model are shown in Fig. 5, with only EPIC-PN data shown for clarity, and the best-fit parameters 
summarized in Table 3.
The low energy spectra below $\sim 0.7$ keV appear flattened in the high flux state, while 
the shape of the spectrum in the high energy part (1--1.7 keV) is relatively similar in both flux states.
If the narrow energy band (0.35--0.7 keV) is approximated by  a power law absorbed by the Galactic column density,
the slopes for the high and low fluxes are $1.5\pm0.5$ and $3.3\pm0.4$, respectively.
The behavior  in the soft ($<$0.7 keV) and hard (1--1.7 keV) bands corresponds to an increase 
in the absorption column density and to a constant inner temperature in
the MCD model fits, respectively. Spectral fits with the compTT model resulted in an extremely large optical depth of the
Comptonizing plasma. In such a situation, more low energy photons are upscattered and the lower part of the spectrum becomes flat.
Fig. 6 compares the best-fit compTT models in the high and low flux states. 

The observed flux in observation 3 is lower than that in observation 2, and we divided the data using a threshold
0.015 counts s$^{-1}$ per PN in the 0.5--1.7 keV band (observed frame).
The spectra are fitted with the same models as in observation 2.
The spectral parameters in the high and low flux intervals are consistent with each other and no significant spectral
change is seen. The temperatures for the MCD and compTT models in the low flux spectrum 
($0.12^{+0.03}_{-0.02}$ and $0.11\pm0.03$ keV, respectively) are slightly lower than those obtained for the averaged spectrum.
The observed flux for observation 1 is even lower, 
and photon statistics of the divided spectra are found to be poor.

\section{Discussion}

\subsection{Is 2XMM~J1231 an AGN?}

The error circle of the X-ray position contains only one 
optical source SDSS~J123103.24 +110648.5 classified as a galaxy and one possible interpretation is that 2XMM~J1231 is an AGN.
We examine the possibility that the detected X-ray source is not an AGN. 
The apparent size of this optical source ($\sim 2$ arcsec) 
is much smaller than the error circle 
and the source could be an off-nuclear source in this galaxy. The observed luminosity $(1.6-3.8)\times10^{42}$ erg s$^{-1}$,
however, is much higher
than ordinary X-ray binaries and ultraluminous X-ray sources (ULXs, e.g., Makishima et al. 2000). The most extreme cases of ULXs show luminosities 
in excess of $10^{42}$ erg s$^{-1}$ and are known as hyperluminous X-ray sources (HLXs). 
The luminosity of HLX-1 in ESO~243-49 reached $1.1\times 10^{42}$ erg s$^{-1}$ in the 0.2--10 keV band (Farrell et al. 2009).
The luminosity of 2XMM~J1231 is even larger than this maximum luminosity of an HLX observed so far. Thus, if 2XMM~J1231 is
an off-nuclear source, it would be the most luminous HLX. The spectra of HLX-1 in ESO~243-49 were described by a steep power law model
or a combination of MCD and power law models depending on the epoch of observation (Farrell et al. 2009).
The spectrum of 2XMM~J1231 is completely different from these spectral shapes, and could be a new state of HLX, if this
object is indeed an HLX.

\begin{figure}[ht]
\begin{center}
\hspace{-1cm}
\includegraphics[angle=-90,scale=0.36]{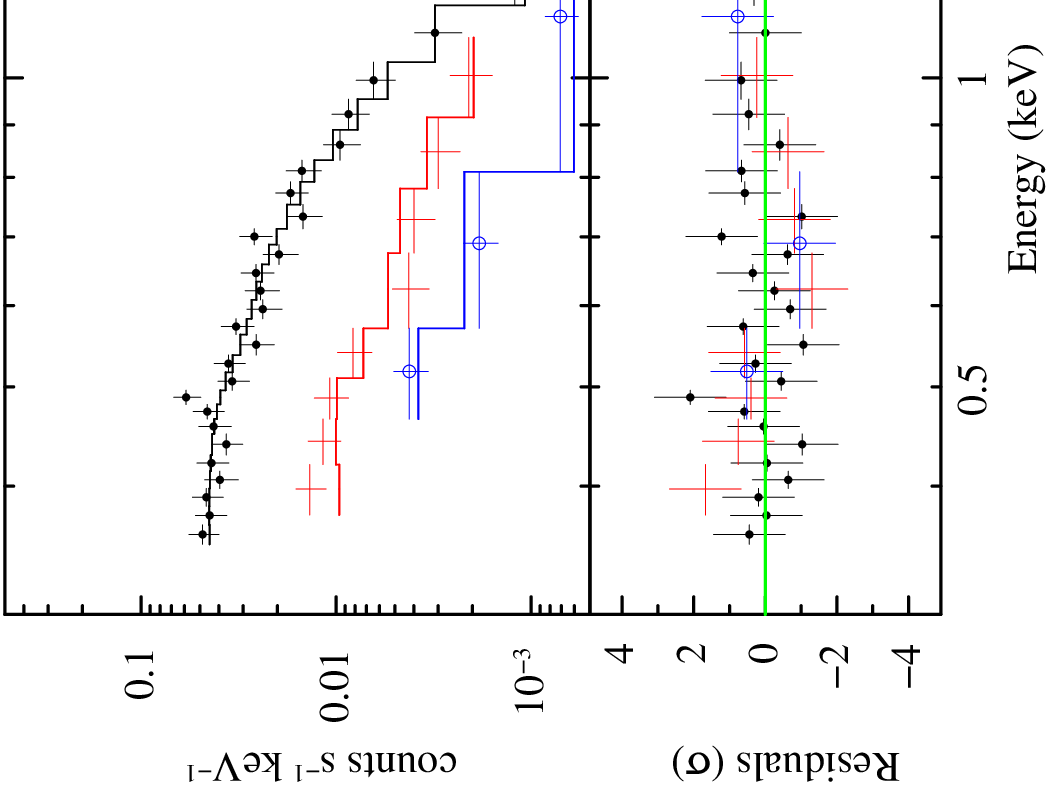}

\hspace{-1cm}
\includegraphics[angle=-90,scale=0.36]{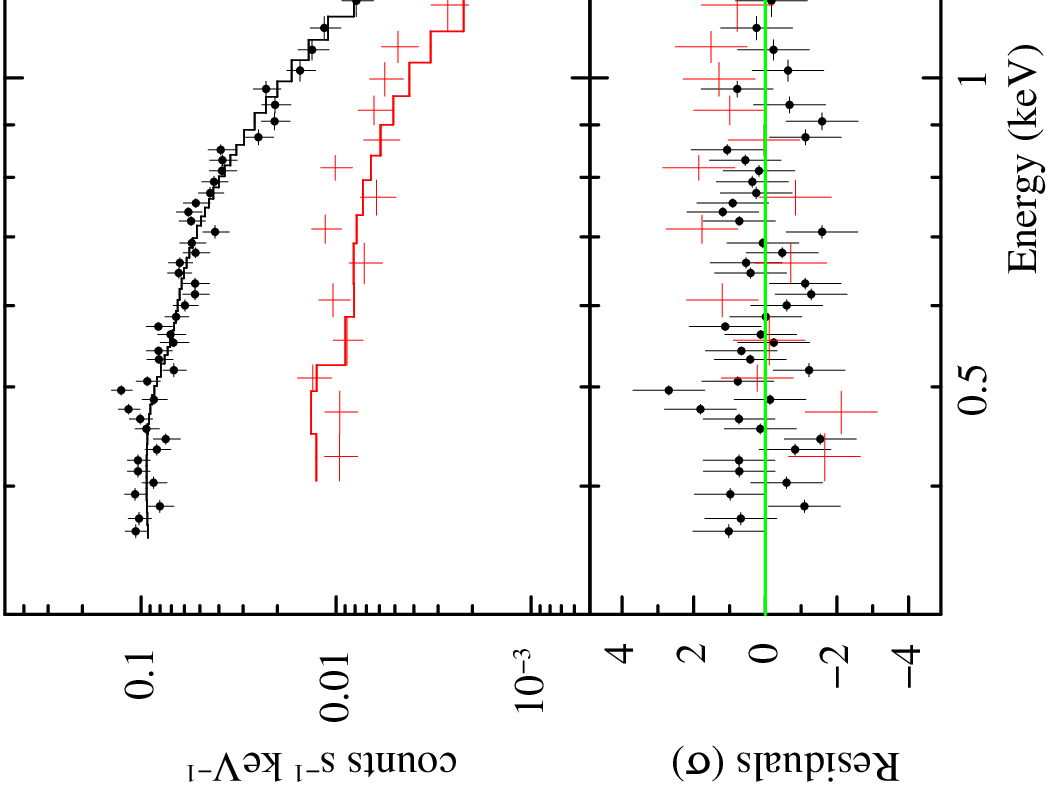}

\hspace{-1cm}
\includegraphics[angle=-90,scale=0.36]{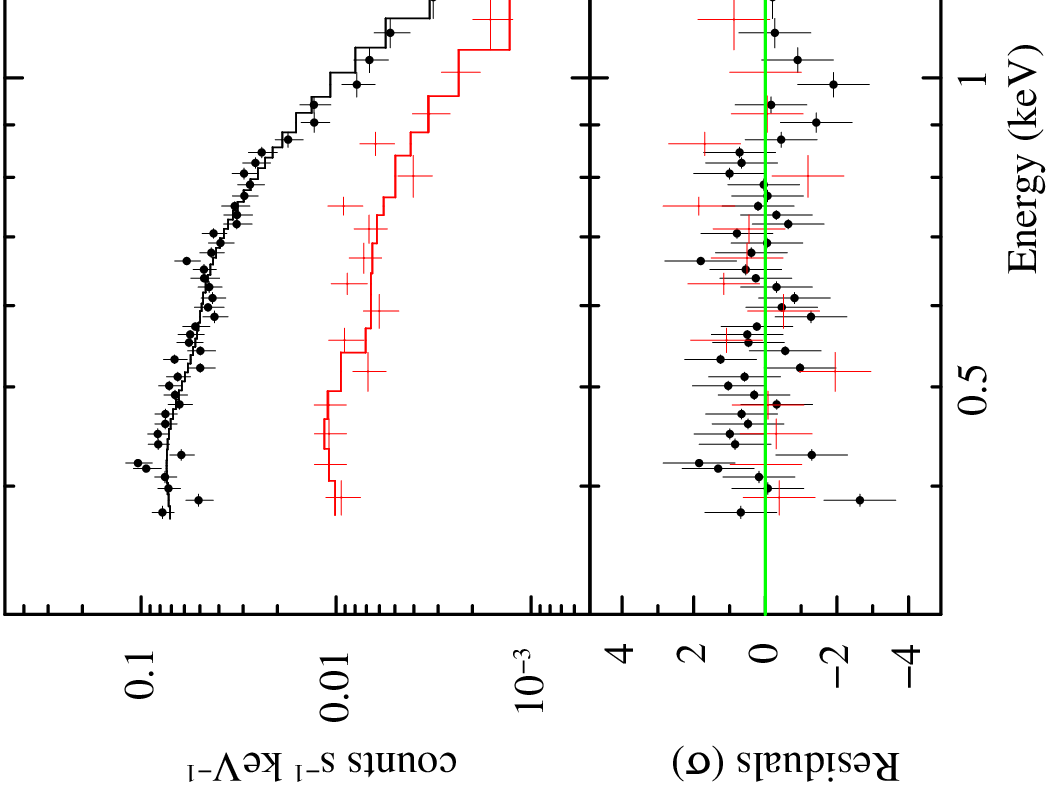}
\caption{EPIC
spectra fitted with a
multicolor disk blackbody model
 for observations 1 to 3.
Lower panel shows residuals.
EPIC-PN, EPIC-MOS1, and EPIC-MOS2 data are plotted as crosses with filled circles, 
crosses with open circles, and crosses without symbols, respectively.
}
\end{center}
\end{figure}

Next we examine the possibility that the association of 2XMM~J1231 with SDSS~J123103.24 +110648.5 is chance coincidence.
The luminosities of 2XMM~J1231 in the 0.5--2 keV band for an assumed distance of 
$d_{\rm 1kpc}$ are $(3.5, 8.8, 8.2)\times 10^{30}d^2_{\rm 1kpc}$ erg s$^{-1}$ for observations 1, 2, and 3, respectively,
where $d_{\rm 1kpc}$ is in units of 1 kpc.  These luminosities correspond to Eddington ratios of
$(2.3-5.9)\times10^{-8} (M_{\odot}/M)$, where $M$ is the mass of the central object.
Because of the high Galactic latitude, it is unlikely that this source is much farther than
the scale height of the Galactic disk.
If so, then the luminosities and Eddington ratios are extremely low compared to Galactic neutron stars (NS) and 
BHBs explored so far (e.g., Asai et al. 1998).
Spectra of BHBs in their quiescent state show power law emission 
(Kong et al. 2002, Hameury et al. 2003)
and are
not similar to what we observed from 2XMM~J1231. Spectra of quiescent NS are often described by
a combination of blackbody and power law, while spectra of some NS are described by a blackbody with
temperature 0.3 keV (Asai et al. 1998, Wijnands et al. 2002), which is somewhat higher than that observed from 
2XMM~J1231. Thus 2XMM~J2131 is less likely to be a Galactic compact source.

The Galactic longitude of 2XMM~J1231 is $73^\circ.3$ and the probability that a relatively bright 
Galactic source coincides with the position of the galaxy appears to be low. We examined the possibility 
of chance association by referring to  previous X-ray surveys. Among soft X-ray surveys, {\it ROSAT} 
International X-ray/Optical Survey (RIXOS), a serendipitous source survey using {\it ROSAT} PSPC 
fields at high Galactic latitude ($|b| > 28^\circ$), has a flux limit of $3\times10^{-14}$ erg cm$^{-2}$ 
s$^{-1}$ in 0.5--2 keV, which is similar to the fluxes observed from 2XMM~J1231
( $(2.5, 7.0, 4.2)\times 10^{-14}$ erg cm$^{-2}$ s$^{-1}$ in 0.5--2 keV for the three observations, respectively). 
344 among 401 X-ray sources detected in RIXOS are optically identified.
238 are identified as galaxies including AGN, emission line galaxies, and normal galaxies, and 78 are stars. 
Thus a large fraction of X-ray sources at this flux level are extragalactic source. 
Mason et al. (2000) studied the relation between X-ray fluxes and 
optical magnitudes and found that  stars are relatively bright in optical; 
all the stellar sources are brighter than 18.5 mag in $R$ band, while there are many fainter extragalactic 
sources down to $R \sim 22$.
The SDSS image around the position of 2XMM~J1231 shows only one optical source 
SDSS~J123103.24+110648.5, which has an $r$ band magnitude of $20.3\pm0.04$.
If this SDSS object is not the counterpart, the true optical counterpart would be fainter than this magnitude
and the optical to X-ray ratio would be extremely small compared to any of the stellar sources found in RIXOS. 
The Galactic extinction toward this object is not large ($A_V = 0.11$ mag) and the optical faintness 
is not due to extinction.
Therefore, we conclude that it is unlikely that the detected source is a Galactic source.

\subsection{A New Spectral State in AGN?}


The spectra of 2XMM~J1231 are very soft and empirically fitted by a
single component model MCD or compTT.
The inner disk temperature obtained by the MCD model $kT =0.16-0.21$
is  within the temperature distribution of the soft excess component seen in AGN (0.16--0.21 keV; 
Gierli\'nski \& Done 2004; Piconcelli et al. 2005; Crummy et al. 2006; 
Bianchi et al. 2009; Ai et al. 2011).
The uniqueness of this source is the dominance of the soft X-ray component. Some AGN show
very strong soft excess relative to a power law component seen at
higher energies. In order to quantify the relative strength of the soft excess, we calculated the ratio of $EF_E$ values at 1 keV and 5 keV
for 2XMM~J1231 and for AGN with strong soft excess, where $E$ is the energy and $F_E$ is the specific energy flux. 
Since hard X-rays were not detected from 2XMM~J2131, the 90\% confidence upper limit on the X-ray counts 
in 2--7 keV was converted to an upper limit on the flux at 5 keV
by assuming that the spectral shape in the hard band is a power law with a photon index of 2.0. 
In these calculations, we combined the two observations with good signal to noise ratios (observations
2 and 3). The $EF_E$ spectra of the best-fit MCD and the upper limit on the hard power law emission are
shown in Fig. 7.
The lower limit on the ratio was calculated to be $>28$. The ratios for the NLS1s
1H0707$-$495 (in its bright state), RE1034+396, RX~J0136.9$-$3510, and IRAS 13224$-$3809, all of which are 
known to show remarkably strong soft excess, are derived to be 3--4, 4, 3, and 3, respectively, from {\xmm} spectra
(Zoghbi et al. 2010, Fabian et al. 2009, Middleton et al. 2009, Jin et al. 2009, Ponti et al. 2010). Thus, 
the soft emission relative to hard X-rays in 2XMM~J1231 is the strongest among AGN.

\begin{figure}[htb]
\begin{center}
\hspace{-2.5cm}
\includegraphics[angle=-90,scale=0.36]{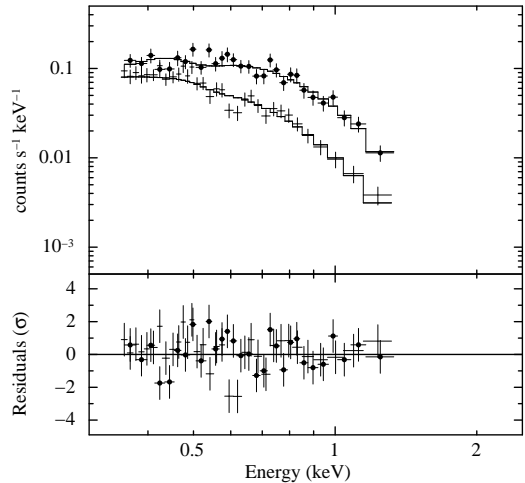}
\end{center}
\caption{Spectra in the high- (upper crosses with filled circles) and low-flux (lower crosses) states
for observation 2
fitted with a multicolor disk blackbody model. Lower panel shows residuals. Only EPIC-PN data are shown for
clarity.
}
\end{figure}

\begin{figure}[ht]
\begin{center}
\includegraphics[angle=-90,scale=0.35]{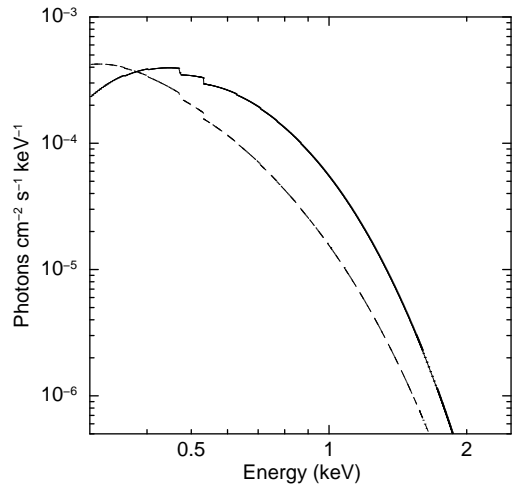}
\caption{Best-fit Comptonization model in high- (solid line) and low-flux
(dashed line) intervals for observation 2.
}
\end{center}
\end{figure}

\begin{figure}[hbt]
\begin{center}
\includegraphics[angle=-90,scale=0.36]{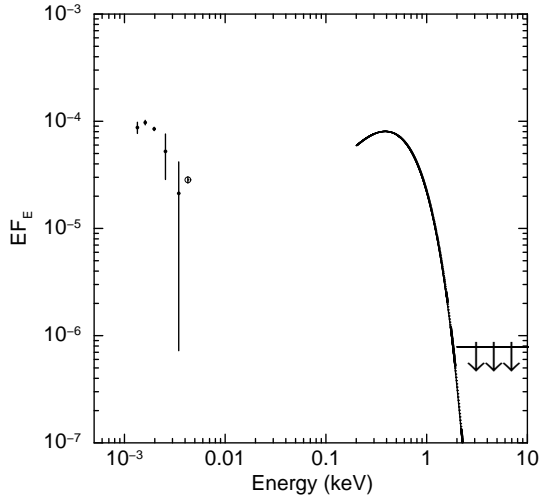}
\end{center}
\caption{
Spectral energy distribution (SED) from the optical to X-ray bands.
Optical data (filled circles) are in the $z$, $i$, $r$, $g$, and $u$ bands obtained with SDSS, and the UV data (open circle) 
are for the UVW1 filter data with OM.
The X-ray SED is shown by the MCD model (solid line) and an upper limit on a power law component (horizontal line with arrows
shown in 2--10 keV) at a 90\%  confidence level derived from a combination of observations 2 and 3.
The upper limit is calculated in the 2--7 keV band assuming a power law with photon index of 2.0 and extrapolated to 10 keV.
The Galactic and intrinsic absorption column densities are set to zero.
}
\end{figure}


The spectral energy distribution (SED) from the optical to UV bands is also shown in Fig. 7, where the optical data
points are PSF magnitudes obtained with the SDSS data in the $u$, $g$, $r$, $i$, and $z$ bands, and the UV data are
the OM data with the UVW1 filter. The spectral shape is falling toward the UV and 
the optical emission
is likely to be dominated by stellar emission in the host galaxy 
as expected from the relatively low luminosity of the candidate AGN. The UV data point at the effective wavelength 
of 2910 \AA \ could also contain stellar emission.  Unless almost all the UV emission comes from stellar components,
the UV to soft X-ray ratio is within a range of those observed from AGN (e.g., Jin et al. 2012). The hard X-ray emission,
however, is extremely weak, and therefore UV to hard X-ray ratio in this object is
unusually large compared to previously known AGN.


2XMM~J1231 showed large amplitude variability in the soft X-ray band. Many highly variable AGN  such as
NLS1s exhibit significant variability in the soft X-rays (e.g., Gallo et al. 2004, 2007; O'Neill et al. 2007; 
Larsson et al. 2008). Such variability is stark contrast to the H/S state of Galactic BHBs
and a small
number of NLS1s (RE~J1034+396, Middleton et al. 2009; RX~J0136.9-3510, Jin et al. 2009), which show relatively 
stable soft X-rays. Thus the X-ray variability and apparent spectral shape in the soft X-ray band 
of  2XMM~J1231 are quite similar to what is seen in many NLS1s, except for the extreme strength of its soft excess emission.
The origin of the soft excess in NLS1 still remains to be understood and several possibilities have been proposed
including soft continuum models and a continuum modified by emission and/or absorption due to atomic processes
(e.g., Middleton et al. 2009).  In the present study, we examined only simple continuum models to represent the soft emission, 
since a broadband continuum should be used to properly test models incorporating atomic features.
One of the continuum models (MCD) seems unlikely as the origin of the highly variable soft emission,
since MCD predicts that the inner temperature increases as the luminosity goes up contrary to what is observed. 
It might, however, still be possible for the spectrum in the low flux state to be of MCD  origin representing 
relatively stable underlying emission, and for the variability to be caused by an additional component such as
strongly Comptonized emission. 
We examine whether this interpretation is 
physically self-consistent by deriving the BH mass and Eddington ratio from the result of the fit. 
If the BH mass is derived from the parameters of 
the MCD fit for the low-flux spectra in observation 2, a BH mass $M_{\rm BH} = 2.8\times10^{4}M_\odot$ is obtained,  
where we assumed a color temperature correction factor of $\kappa = 1.7$, the correction for the fact that the actual radius of maximum temperature is larger than the disk inner radius $\xi = 0.41$ (see Kubota et al. 1998 for derivation), and the disk inner radius equals $3R_{\rm S}$
(i.e., innermost stable circular orbit for a non-spinning BH).
Then the disk bolometric luminosity $L_{\rm disk} = 5.9\times10^{42}$ erg s$^{-1}$ corresponds to an
Eddington ratio of 1.4.
An analogy with BHBs suggests that bare disk emission is not directly seen because of substantial inverse Compton scattering by coronae at such a high Eddington ratio
(``very high state"\ in BHB, e.g., Kubota et al. 2001; Kubota \& Done 2004; Done et al. 2007). 
Thus the MCD interpretation appears unviable even for the low-flux state.

The relative strength of the soft excess in X-ray spectra of AGN depends on the Eddington ratio; objects with
a large Eddington ratio tend to show stronger soft excess relative to hard X-ray emission (Jin et al. 2012).
Assuming this trend holds in 2XMM~J1231, the extreme soft excess implies that its Eddington ratio is very large.
Since the SED of this object is not well constrained from the data,
we assume that the disk luminosity derived from
the MCD fit ($L_{\rm disk} = 6.7\times10^{42}$ erg s$^{-1}$) approximates the bolometric luminosity
$L_{\rm bol}$. Then the Eddington ratio is written as
\[
L_{\rm bol}/L_{\rm Edd} = 0.45 \left( \frac{10^5 M_\odot}{M_{\rm BH}}  \right).
\]
In order to achieve an Eddington ratio greater than $\sim 0.3$, for which X-ray spectra of some BHBs exhibit 
signature of Comptonization (Done et al. 2007), the BH mass should be smaller than $1.5\times10^{5}M_\odot$.
This BH mass is the smallest among AGN (Greene \& Ho 2004, 2007a, 2007b; Ai et al. 2011).

Soft X-ray emission suffering from Comptonization by plasma with a large optical depth has been reported for 
ULXs and the microquasar GRS~1915+105. Such a state is referred to as the ultraluminous state  
(Roberts 2007, Gladstone et al. 2009, Vierdayanti et al. 2010b) and appears at a luminosity 
near or above Eddington. 
In our fits using the compTT model,
the seed photons are assumed to be from an accretion disk with a temperature of 0.01 keV as expected for the inner part of accretion disks around super massive BHs. 
The value of the optical depth of scattering plasma ($\tau \approx 21-30$ for the averaged spectra) is extremely
large compared to that observed in GRS~1915+105 ($\tau < 10$, e.g., Ueda et al. 2009, Vierdayanti et al. 2010b). Because of the low temperature of the underlying disk, only emission  Comptonized by optically thick plasma is seen in the observed bandpass, and therefore the optical depths obtained from our fits could have large uncertainties  and should be interpreted with caution.
The Comptonized emission in BHBs and ULXs could show variability (such as low frequency quasi-periodic oscillations 
in GRS 1915+105, e.g., Ueda et al. 2010;  Vierdayanti et al. 2010a) presumably caused by  changes in physical parameters of Comptonizing clouds. Similar 
mechanisms could be at work in 2XMM~J1231. Alternatively, variability could be due to instabilities inside the disk itself as seen in some states in GRS1915+105 showing limit cycle oscillations on timescales of 10--100 s (Belloni et al. 2000). In such states, the Eddington ratio is large (0.3--3) and X-ray spectra show strong optically thick Comptonization (Zdziarski et al. 2005, Ueda et al. 2010).  Future long observations will be able to test whether
the variability observed in 2XMM~J1231 really shows limit-cycle oscillations, which has never observed in AGN, and whether spectral variability is accompanied by such oscillations.

\acknowledgements

The authors thank Poshak Gandhi for his critical reading of the manuscript 
and the referee for useful suggestions that improved this paper.
This work was partly supported by the Grant-in-Aid for Scientific Research
20740109 (Y.T.), 21244017 (H.A.), and 23540265 (Y.U.) 
 from the Ministry of Education, Culture, Sports, Science,
and Technology of Japan.

{\it Facility}: XMM


\end{document}